\documentclass[iop,apj]{emulateapj}
\usepackage[dvips]{epsfig}
\usepackage{float}
\usepackage{amsmath}
\usepackage{graphicx}
\def \be{\begin{equation}}
\def \ee{\end{equation}}
\def \bdm{\begin{eqnarray}}
\def \edm{\end{eqnarray}}
\begin{document}
\title{Theoretical Explanation of the Cosmic Ray Perpendicular Diffusion Coefficient in the Nearby Starburst Galaxy NGC 253}
\author{K. Buffie$^{1}$, V. Heesen$^{2}$, and A. Shalchi$^{1}$}
\affil{$^{1}$Department of Physics and Astronomy, University of Manitoba,
  Winnipeg, Manitoba R3T 2N2, Canada,
 shalchi@physics.umanitoba.ca}
\affil{$^{2}$School for Physics and Astronomy, University of Southampton, Southampton SO17 1BJ, UK, v.heesen@soton.ac.uk}
\begin{abstract}
Diffusion coefficients are usually used to describe the propagation of Cosmic Rays through the Universe.
Whereas such transport parameters can be obtained from experiments in the Solar System, it is difficult
to determine diffusion coefficients in the Milky Way or in external galaxies. Recently a value for the
perpendicular diffusion coefficient in the nearby starburst halaxy NGC\,253 has been proposed. In the
present paper we reproduce this value theoretically by using an advanced analytical theory for perpendicular
diffusion.
\end{abstract}
\section{Introduction}
Radio continuum observations of the nearby starburst galaxy NGC\,253 can be used
to measure the distribution and transport of Cosmic Ray electrons \citep[see][]{Heesen09a,Heesen11}.
The current understanding is that Cosmic Ray electrons are accelerated in the disk by shock waves induced
by supernova explosions \citep{Reynolds12}. The electrons are then transported away over their lifetime by either convection
in a galactic wind or diffusion, where the two processes can be distinguished from their different typical
transport length as a function of electron energy. The magnetic field in galaxies is normally dominated
by the turbulent component, but averaged over spatial sizes larger than 1\,kpc\footnote{In the present
article we measure distances in parsecs (pc). Approximately we have $1$pc $\approx 3 \cdot 10^{16}$m
$\approx 3.3$ly.} (a typical resolution achieved with radio interferometers in external galaxies) ordered
magnetic field components are detected by their linear polarisation from synchroton emission of Cosmic Ray
electrons spiralling in the magnetic field. Observations suggest that galaxies in their disk plane are
dominated by a magnetic field that is parallel to the disk and has a spiral pattern with a constant
pitch angle of a few tens of degrees \citep{Beck12}. Further away from the disk plane at heights larger
than 1 kpc the field becomes more vertical and the field lines open up further into the halo \citep{Haverkorn12}.

The motion of Cosmic Rays through the interplanetary or interstellar space is complicated due to their
interaction with turbulent magnetic fields. Such fields lead to spatial diffusion of the energetic
particles. In addition to the turbulent magnetic fields one can also observe a non-vanishing mean magnetic
field which is in the case of the interstellar medium the Galactic magnetic field. This mean field breaks
the symmetry of the physical system and one has to distinguish between diffusion along and across the
Galactic magnetic field. The former process is also known as parallel diffusion whereas the latter process
is called perpendicular transport. The theoretical investigation of Cosmic Ray transport has a long
history \citep[see, e.g.,][]{Schlickeiser02}. Whereas the propagation of cosmic particles through the solar
system seems to be well understood \citep[see, e.g.,][]{Shalchi06,Tautz12}, it is unclear
whether advanced transport theories can describe particle propagation with high accuracy in other systems
like the Milky Way or external galaxies. At least parallel diffusion seems to be a process which can be
described very well by more advanced diffusion theories. The measured decrease of the abundance ratio
of secondary to primary Cosmic Ray nuclei as B/C and N/O at kinetic energies above $1$ GeV/nucleon,
implies a variation of the total column density as a function of rigidity $R$ as (Swordy et al. 1990)
$\lambda_{\parallel} \propto R^{0.6}$ where we have used the parallel mean free path $\lambda_{\parallel}$.
This behavior was explained by \citet{Shalchi05} who used a nonlinear diffusion theory.
The behavior of low energy Cosmic Rays was also explained by using an extension of quasilinear
theory \citep[see][]{Shalchi10c}.

More complicated, however, is the process of perpendicular diffusion. In the recent years progress has
been achieved. \citet{Shalchi10a} developed an advanced theory for perpendicular diffusion which contains
all known limits of diffusion theory. This theory is called the Unified Non-Linear Transport (UNLT) theory
and provides the correct subdiffusive behavior for one-dimensional turbulence in agreement with the so-called
theorem on reduced dimensionality (see Jokipii et al.\ 1993 and Jones et al.\ 1998) and computer simulations
\citep[see, e.g.,][]{Qin02}. Furthermore, the correct field line random walk limit can be derived in agreement
with the nonlinear field line transport theory of \citet{Matthaeus95}. The theory discussed here was also
tested numerically confirming its validity for solar wind parameters \citep[see][]{Tautz11}.

\begin{figure}[hbtp]
\begin{center}
\resizebox{0.9\columnwidth}{!}{\includegraphics{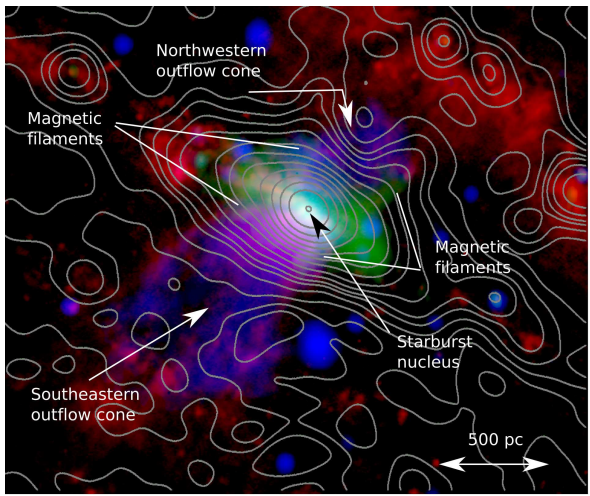}}
\end{center}
\caption{Three-colour composite, multi-wavelength view of the central region in NGC\,253. Red, green, and
blue indicate H$\alpha$ from \citet{Westmoquette11}, $\lambda$20\,cm radio continuum, and \emph{Chandra} soft
X-ray (M. Hardcastle, priv.\ com.), respectively. Contours show $\lambda$3\,cm radio continuum at 150\, pc resolution with the first contour at 3 $\times$ the r.m.s\ noise level and each consecutive contour increasing roughly by a factor of two. An identical figure without the annotations and contours can be found in \citet{Heesen11}.}
\label{fig:cone}
\end{figure}

In analytical theories for energetic particle transport, the model of interstellar turbulence enters the
corresponding nonlinear integral equation for the perpendicular diffusion coefficient. \citet{Shalchi10b}
have therefore combined the UNLT theory with the turbulence correlation tensor proposed by \citet{Cho02}
based on Goldreich-Sridhar scaling \citep[see][]{Goldreich95}. Later, however, \citet{Heesen11}
compared the analytical result obtained in the aforementioned paper with observations of NGC\,253 and noticed
a major difference. Using a multi-wavelength picture of the starburst nucleus they established that the
nuclear outflow of hot X-ray emitting gas can be collimated by the magnetic field lines in the walls that
surround the conical outflow. The magnetic field lines are aligned with the wall orientation as would be
expected from compression by the hot gas in the outflow cone. An annotated figure of the multi-wavelength
picture can be found in Figure~\ref{fig:cone}, which shows the approximate location of the magnetic field
filaments that are the walls of the outflow cone seen in projection. {\bf We have overlaid contours of the $\lambda$3\,cm radio continuum emission at a resolution of 150\,pc that show the extent of the filaments extending along the northwestern outflow cone. The filaments are also observed at $\lambda\lambda$ 20 and 6\,cm at the same spatial resolution, which offers to study the perpendicular diffusion coefficient as described in Section~\ref{sec:observations}.}

{\bf These observations led \citet{Heesen11} to the finding that the observed perpendicular diffusion coefficient is a factor of 10 too high in comparison to the predicted value by the theory.} In the
present paper we, therefore, revisit the problem of perpendicular transport of energetic particles
in the nearby starbust galaxy NGC\,253. We combine the approach of \citet{Shalchi10b} with more realistic
particle and turbulence parameters to compute the perpendicular diffusion coefficient more accurately.
\section{Theoretical Description of Perpendicular Diffusion}
The analytical description of diffusion across the mean magnetic field has a long history \citep[see][for
a review]{Shalchi09book}. Very recently an advanced theory for perpendicular transport has been developed \citep[see][]{Shalchi10a}
which shows good agreement with computer simulations performed for the parameters of the solar wind
\citep[see][]{Tautz11}. This theory is called the Unified Non-Linear Transport (UNLT) theory and provides
a nonlinear integral equation for the perpendicular diffusion coefficient $\kappa_{\perp}$. According to this
equation $\kappa_{\perp}$ depends on magnetic turbulence described by the so-called magnetic correlation tensor.
Therefore, the understanding of turbulence is crucial for the analytical description of cross field diffusion.
A model for the magnetic correlation tensor in the interstellar medium was proposed by \citet{Cho02}
which is based on the Goldreich-Sridhar model \citep{Goldreich95}. For this specific turbulence model,
\citet{Shalchi10b} derived the following equation
\bdm
\frac{\lambda_{\perp}}{\lambda_{\parallel}}
& = & \frac{a^2}{6}\left(\frac{\delta{B}}{B_0}\right)^{2/3}\int_0^1dy\int_0^{\infty}dxe^{-(xy^{2/3}E_B^{-4/3})} \nonumber\\
& \times & \frac{2x^2y^2+1}{x^2y^2+1}
\frac{y^{7/3}}{x^2y^4\lambda_{\parallel}/\lambda_{\perp}+y^2+4\lambda_{\parallel}\lambda_{\perp}/(3l)^2}
\label{diffusion}
\edm
where we have used the perpendicular mean free path which is related to the perpendicular diffusion coefficient via
$\lambda_{\perp}=3 \kappa_{\perp}/3$. According to Eq.~(\ref{diffusion}) the parameter $\lambda_{\perp}$ depends on
the parallel mean free path $\lambda_{\parallel}=3\kappa_{\parallel}/v$ ($v$ is the particle speed) and, therefore,
there is a mutual influence of parallel and perpendicular transport. The parameter $a^2$ in Eq. (\ref{diffusion})
is an order one constant related to the probability that the particle is scattered away from the magnetic field
lines \citep[see, e.g.,][]{Dosch09}. In the present paper we set $a^2 = 1$ for simplicity. The perpendicular diffusion
coefficient is also controlled by the characteristic length scale of the turbulence $l$, and the ratio $E_B \equiv \delta B / B_0$.
The latter ratio is a measure for the turbulent magnetic field strength $\delta B$ with respect to the mean field $B_0$.
The mean field is $B_0$ defined as the spatially averaged magnetic field vector on a scale corresponding to the length $l$,
which is usually related to the correlation length of magnetic turbulence. The latter parameter is a characteristic scale
for the correlation of magnetic fields at two different positions in space. Often this scale is directly proportional to
the so-called bendover scale which denotes the turnover from the inertial range of the turbulence to the energy range.
Furthermore, this scale is smaller than the largest scales on which turbulence can be observed.

\citet{Shalchi10b} have solved Eq.~(\ref{diffusion}) numerically to obtain values for the perpendicular diffusion coefficient
in the interstellar medium. \citet{Heesen11}, however, compared the theoretical results with observations and found a disagreement.
Below we, therefore, perform a parameter study to compute the perpendicular diffusion coefficient and compare it with observations.
We will show that for the correct choice of input parameters, Eq.~(\ref{diffusion}) is indeed able to reproduce the aforementioned
observations.
\section{Observations}
\label{sec:observations}
The different parameters entering Eq.~(\ref{diffusion}) as well is the perpendicular diffusion coefficient itself
can be obtained from radio continuum observations of the nearby galaxy NGC 253 as described in \citet{Heesen09a,Heesen11}.
In this section, we briefly summarize the techniques they have been using to quantify the Cosmic Ray transport. 

Radio continuum studies can be used to measure the length scale of Cosmic Ray diffusion. The time scale of Cosmic Ray electrons is determined by loss processes by which the electrons are losing their energy, such as synchrotron and inverse Compton radiation. The diffusion coefficient can be then calculated by
\begin{equation}
\kappa_{\rm obs} = L_{\rm diff}^2 /\tau,
\end{equation}
where $L_{\rm diff}$ is the diffusion length scale and $\tau$ is the electron lifetime. Depending on the magnetic field structure,
the observed diffusion coefficient is either along ($\kappa_{\parallel}$) the magnetic field or perpendicular to it ($\kappa_{\perp}$).
Ideally, one can measure the distance to the star-formation sites, where the Cosmic Rays are accelerated and injected into the interstellar medium, to obtain the transport length scale. This is for instance the case when observing galaxies in so-called edge-on geometry, where the observer is located in or close to the disk plane of the galaxy (the inclination angle is close to $90^\circ$). Cosmic Ray acceleration in supernova remnants is confined to the relatively thin disc plane, where the formation of massive stars happens. The geometry in this case is thus simple: Cosmic Rays are transported away from the star formation sites over their lifetime and their transport length scale is equal to the vertical electron scaleheight. Typical scaleheights of galaxies are 1.8\,kpc at observing wavelengths of $\lambda$6\,cm \citep{Krause09}, where longer wavelengths have larger scaleheights and shorter wavelengths smaller ones. Typical Cosmic Ray electron lifetimes in the interstellar medium are between $1$ and $10$\,Myr. This basically determines the typical parallel diffusion coefficient in the ISM, where it is assumed that the halo magnetic field is opening up from the disk parallel to a vertical direction with increasing distance from the disk. Therefore, the vertical diffusion is predominantly along the magnetic field lines hence allowing us to measure $\kappa_{\parallel}$. \citet{Heesen09b} confirmed this for NGC\,253 finding the characteristic X-shaped halo magnetic field with significant vertical components that is observed in a number of nearby galaxies \citep[e.g.,][]{Soida11}.

To measure the perpendicular diffusion coefficient is more difficult, because it requires to have variations of the Cosmic Ray distribution perpendicular to the magnetic field orientation. The observed radio continuum emission is mostly smooth in the disk-halo interface and filamentary structures are rare. Numerical MHD simulations suggest that the disk halo interface is dominated by filamentary magnetic fields \citep{Breitschwerdt12}, but line-of-sight confusion and limited spatial resolution hampers their detection. An exception are starburst galaxies, where the spatially concentrated star formation activity results in exceptionally high radio continuum surface brightness, allowing us to employ high spatial resolution ($\approx 100\,\rm pc$). The spatially concentrated star formation can result into outflows of hot X-ray emitting gas in a galactic wind. The magnetic field is then concentrated and amplified by expansion of the hot gas until a pressure equilibrium is reached. This very specific geometry allowed \citet{Heesen11} to measure the perpendicular diffusion across the magnetic field in the walls of the nuclear outflow cone in NGC\,253. 

The latter authors found the width of radio continuum filaments weakly dependent on frequency and thus
electron age, which can be interpreted as perpendicular Cosmic Ray diffusion. We note that they were
able to measure the orientation of the ordered magnetic field along the filaments as it would be
expected for a compression due to the hot X-ray emitting gas outflowing from the nuclear starburst.
From the observations of particles with a magnetic rigidity of $R=3 \times 10^{12}$ Volt (equivalent to an electron energy of 3\,GeV), they found a perpendicular diffusion coefficient of $\kappa_{\perp}=(2.6\pm 0.6) \times 10^{28}\,\rm cm^2\,s^{-1}$. For the parallel
diffusion coefficient one can assume $\kappa_{\parallel} = 1.0 \times 10^{29}\,\rm cm^2\,s^{-1}$ (again for an electron energy of 3\,GeV), as measured from the diffusion of Cosmic Ray electrons from the disk into the halo along the vertical halo magnetic field \citep{Heesen09a}. It is important to measure the diffusion coefficients at roughly the same electron energy, because they are energy dependent. We have reduced the measured diffusion coefficient of \citet{Heesen09a} to account for a possible contribution of convection. The southwestern halo is different from the northeastern one as indicated by the dependence of the electron scaleheight on the electron lifetime and by the different amount of extra-planar gas that is more abundant in the northeastern convective halo. However, we can not rule out the contribution of convection of Cosmic Rays by the disc wind and thus have lowered the diffusion coefficient assuming that both convection and diffusion contribute equally in the southwestern halo.

Equation~(\ref{diffusion}) depends on turbulence properties, namely the ratio of magnetic fields $\delta B/B_{0}$ and
the scale $l$. According to \citet{Heesen11} the equipartion estimate for the total field strength is $B_{\rm tot}=46\,\mu\rm G$.
The ordered field strength $B_{\rm ord,\perp}$ in the sky plane can be computed from the observed degree of polarization $p$
of the synchrotron emission \citep[e.g.,][]{Beck05}:
\begin{equation}
p = p_0 \mbox{${\left( 1 + \frac{7}{3} q^2 \right)}$/${\left( 1 + 3q^2 + \frac{10}{9} q^4 \right)}$},
\end{equation}
where $p_0$ is the intrinsic degree of polarization ($p_0=(3-3\alpha)/(5-3\alpha)$) and $q$ is the ratio of the isotropic
turbulent field $B_{\rm turb}$ to the ordered field  $B_{\rm ord,\perp}$ in the plane of the sky. For a non-thermal radio
spectral index of $\alpha=-1$ and a polarization degree of $p=0.21$ we obtain $B_{\rm turb}=41\,\rm \mu G$ and $B_{\rm ord,\perp} = 21\,\rm \mu G$.
For this particular case, the ordered magnetic field lies in the sky plane so that we find for the ordered magnetic field strength
$B_0= 21\,\rm \mu G$. The fluctuations in the magnetic field strength are equal to the turbulent field field strength
($\delta B = B_{\rm turb}$) and thus we find for the the ratio $\delta B / B_0 \approx 2$. We note that this is an upper limit as our spatial resolution of $150\,\rm pc$ that was available for the polarization measurements may be not be enough to resolve the filamentary magnetic fields. High resolution $\lambda$ 20\,cm maps show a filament width of only 40\,pc  in radio continuum emission (see Figure~\ref{fig:cone}), so that our measured ordered magnetic field strength $B_{\rm ord}$ is only a lower limit. We have thus studied the implication of a lower value of $\delta B / B_0 = 1$ to indicate how it would change the theoretical expectation for the perpendicular diffusion coefficient.
%
%
%
%
%
%

More difficult to estimate is the scale $l$. The width of the cone walls in which the magnetic fields are confined is equal
or less than 40\,pc \citep{Heesen11}. This suggests that the upper value cannot be larger than about  $l \approx 50-100$ pc.
However the value for $l$ is very uncertain and, therefore, we compute the perpendicular diffusion coefficient for a whole
range of correlation lengths. Beck (2007) for instance suggested that the largest scales of turbulence are in the order
of $10-100\,\rm pc$. Such largest scales, however, are not necessarily equal to the turbulence correlation scale. Actually they
can be seen as the maximum of the scale $l$. Sometimes \citep[see, e.g.,][]{Shalchi09} it is assumed that the
correlation length of interstellar turbulence is $1$ pc but it could also be shorter. Therefore we compute
the perpendicular diffusion coefficient for $1\,{\rm pc} \leq l \leq 1000\,{\rm pc}$ to explore the values of $l$ which lead
to agreement between theory and observations.

%
%
%
%
%
\begin{figure}[hbtp]
\begin{center}
\resizebox{0.9\columnwidth}{!}{\includegraphics{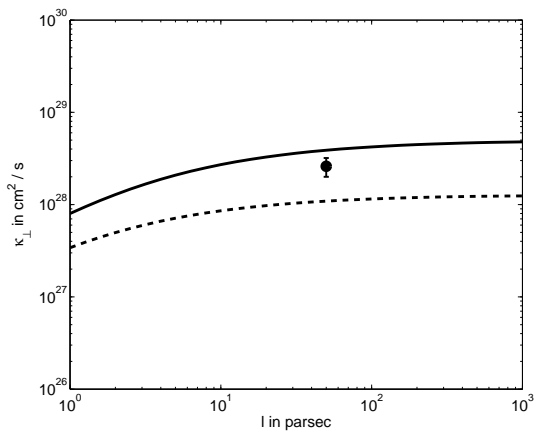}}
\end{center}
\caption{The perpendicular diffusion coefficients of Cosmic Rays in the nearby starbust galaxy NGC\,253. We show
the theoretical perpendicular diffusion coefficient for different length scales $l$. For the parallel diffusion
coefficient we have used $\kappa_{\parallel}=1.0 \times 10^{29}\,\rm cm^2\,s^{-1}$ as proposed by \citet{Heesen09a}.
The theoretical values were calculated for $\delta B / B_0 =1$ (dashed line) and $\delta B / B_0 =2$ (solid line).
The dot represents diffusion coefficient from the observations \citep[see][]{Heesen11}, where we have
$\kappa_{\perp}=(2.6\pm 0.6) \times 10^{28}\,\rm cm^2\,s^{-1}$ at $l=50\,\rm pc$.}
\label{ISMperpf2ref}
\end{figure}
\section{Reproducing the Observations}
In the following we solve Eq. (\ref{diffusion}) numerically for different turbulence scales $l$. To relate
the mean free paths to the spatial diffusion coefficients we employ $\lambda_{\parallel}=3 \kappa_{\parallel}/v$
and $\lambda_{\perp}=3 \kappa_{\perp}/v$. In the latter relations we replace the particle speed $v$ by the
speed of light $c$ since we deal with relativistic particles. As described in the previous sections we set
$\kappa_{\parallel} = 1.0 \times 10^{29}\,\rm cm^2\,s^{-1}$ and $a^2 =1$ in our equation for the perpendicular
diffusion coefficient. We have calculated the latter transport parameter for $\delta B / B_0 = 1$ and
$\delta B / B_0 = 2$. Our theoretical results are shown together with the observations in Fig. \ref{ISMperpf2ref}.

The observational value of the perpendicular diffusion coefficient is $(2.6\pm 0.6) \times 10^{28}\,\rm cm^2\,s^{-1}$ for a correlation
length of approximately 50\,pc. The theoretical perpendicular diffusion coefficient is calculated for different values
of the correlation length. For a correlation length of 50\,pc we found a $\kappa_{\perp}$
which is very close to the observations depending of the value of $\delta B / B_0$. We note that the perpendicular diffusion
coefficient is only weakly dependent on the turbulent correlation scale above a scale of approximately 10\,pc. The observed
diffusion coefficient seems to rule out correlation lengths smaller than 1\,pc, but does not constrain very well the upper
limit. It is reasonable to assume that the correlation length is smaller than the largest scales which are in the order
of $10-100\,\rm pc$ \citep[see][]{Beck07}. We note that the observed value of the perpendicular diffusion coefficient is quite well reproduced by the measured value for $\delta B / B_0$, which is in the range between 1 and 2 as argued in Section~\ref{sec:observations}.
%
%
%
%
%
%
%
%
%
\section{Summary and conclusion}
A fundamental process in astrophysics is the propagation of Cosmic Rays through the Universe. Whereas it seems
that we understand the motion of energetic particles in the solar system very well \citep[see, e.g.,][]{Shalchi06,Tautz12}, it is still not clear how the transport parameters look like in the interstellar space
of our own or external galaxies. \citet{Shalchi10b} have calculated the perpendicular diffusion coefficient
for interstellar turbulence parameters. \citet{Heesen09a, Heesen11} obtained measurements for the parallel and
perpendicular diffusion coefficient in the nearby starbust galaxy NGC\,253. They compared their results with the
theoretical values presented in \citet{Shalchi10a} and found one order of magnitude between the two results.
However, the perpendicular diffusion coefficient obtained in \citet{Shalchi10b} was obtained for a different
set of parameters. Especially the parallel diffusion coefficient used in the latter paper does not agree with
the value obtained by the observers. Thus, in the current paper we combined the approach of \citet{Shalchi10b}
with parameter sets tailored to the observations of NGC\, 253, such as the parallel diffusion coefficient and
magnetic fields strengths. Furthermore, we computed the parameter $\kappa_{\perp}$ for different values of the correlation
length $l$ as this value is known only to a small degree of certainty. Our results now largely reconcile the
observed values with the theoretical predicted ones as visualized in Fig. \ref{ISMperpf2ref}. For correlation
lengths between $5$ and 100\,pc, the theoretical value and the observed one agree within a factor of $2$.
This agreement is quite good given the difficulties in the observational measurements of diffusion coefficients for
Cosmic Ray electrons from radio continuum observations in galaxies. We can confidentially exclude correlation lengths
smaller than $1$\,pc, which would result in too small perpendicular diffusion coefficients.

We have done this study for only one galaxy, because so far measurements particularly of the vertical diffusion
coefficient are very sparse in galaxies. However, the agreement of the observations with the theoretical
prediction lends to some degree support both to the theoretical description of Cosmic Ray diffusion as described
in this paper and to the observational attempts to measure it from the Cosmic Ray electron distribution.
\begin{acknowledgments}
{\it Support by the Natural Sciences and Engineering Research Council (NSERC) of Canada is acknowledged.
V.H. is funded as postdoctoral research assistant by the UK's Science and Technology Facilities Council (STFC).}
\end{acknowledgments}

\end{document}